\documentclass[aps, prx, reprint, superscriptaddress]{revtex4-2}

\usepackage{amsmath}
\usepackage{amssymb}
\usepackage{wasysym}
\usepackage{graphicx}
\usepackage{hyperref}
\usepackage{color}
\usepackage{physics}
\usepackage{siunitx}
\usepackage{xcolor}
\usepackage{changes}

\allowdisplaybreaks

\begin{document}

\title{Collisions of ultracold molecules in bright and dark optical dipole traps}

\author{Roman~Bause}
\author{Andreas~Schindewolf}
\author{Renhao~Tao}
\author{Marcel~Duda}
\author{Xing-Yan~Chen}
\affiliation{Max-Planck-Institut f\"{u}r Quantenoptik, 85748 Garching, Germany}
\affiliation{Munich Center for Quantum Science and Technology, 80799 M\"{u}nchen, Germany}
\author{Goulven Qu\'{e}m\'{e}ner}
\affiliation{Universit\'{e} Paris-Saclay, CNRS, Laboratoire Aim\'{e} Cotton, 91405 Orsay, France}
\author{Tijs~Karman}
\affiliation{Radboud University, Institute for Molecules and Materials, Heijendaalseweg 135, 6525 AJ Nijmegen, The Netherlands}
\author {Arthur~Christianen}
\affiliation{Max-Planck-Institut f\"{u}r Quantenoptik, 85748 Garching, Germany}
\affiliation{Munich Center for Quantum Science and Technology, 80799 M\"{u}nchen, Germany}
\author{Immanuel~Bloch}
\affiliation{Max-Planck-Institut f\"{u}r Quantenoptik, 85748 Garching, Germany}
\affiliation{Munich Center for Quantum Science and Technology, 80799 M\"{u}nchen, Germany}
\affiliation{Fakult\"{a}t f\"{u}r Physik, Ludwig-Maximilians-Universit\"{a}t, 80799 M\"{u}nchen, Germany}
\author{Xin-Yu~Luo}
\email{xinyu.luo@mpq.mpg.de}
\affiliation{Max-Planck-Institut f\"{u}r Quantenoptik, 85748 Garching, Germany}
\affiliation{Munich Center for Quantum Science and Technology, 80799 M\"{u}nchen, Germany}

\date{\today}

\begin{abstract}
Understanding collisions between ultracold molecules is crucial for making stable molecular quantum gases and harnessing their rich internal degrees of freedom for quantum engineering. Transient complexes can strongly influence collisional physics, but in the ultracold regime, key aspects of their behavior have remained unknown. To explain experimentally observed loss of ground-state molecules from optical dipole traps, it was recently proposed that molecular complexes can be lost due to photoexcitation. By trapping molecules in a repulsive box potential using laser light near a narrow molecular transition, we are able to test this hypothesis with light intensities three orders of magnitude lower than what is typical in red-detuned dipole traps. This allows us to investigate light-induced collisional loss in a gas of nonreactive fermionic $^{23}$Na$^{40}$K molecules. Even for the lowest intensities available in our experiment, our results are consistent with universal loss, meaning unit loss probability inside the short-range interaction potential.
Our findings disagree by at least two orders of magnitude with latest theoretical predictions, showing that crucial aspects of molecular collisions are not yet understood, and provide a benchmark for the development of new theories.
\end{abstract}

\maketitle
\section{Introduction}
Diatomic molecules lie at an intriguing boundary in our current understanding of quantum physics. They are much more complex and richer objects than atoms, but at the same time simple enough that achieving complete understanding and control at the quantum level appears to be within reach. For this reason, ultracold diatomic molecules have been studied extensively during the last two decades, and have been suggested as candidate systems for quantum simulation of dipolar systems~\cite{Baranov_2012, Matveeva_2012, Yao_2013, Schuster_2019b}, universal quantum computation~\cite{DeMille_2002, Rabl_2006, Ni_2018, Hughes_2020}, and as ultrasensitive probes of beyond-standard-model physics~\cite{Safronova_2018, ACME_2018, Cairncross_2019}. 

Dipolar bialkali dimers are a specifically interesting class of molecules, both because they can be associated from ultracold atoms, and because they offer large permanent dipole moments, making them an ideal choice for the study of dipolar quantum many-body systems. For this reason, a variety of ground-state bialkali dimers have been experimentally investigated, including the fermionic species $^{40}$K$^{87}$Rb and $^{23}$Na$^{40}$K~\cite{Ni_2008, Park_2015a}, as well as the bosonic species $^{87}$Rb$^{133}$Cs, $^{23}$Na$^{87}$Rb, $^{23}$Na$^{39}$K, and $^{23}$Na$^{133}$Cs~\cite{Takekoshi_2014, Molony_2014, Guo_2016, Voges_2020, Cairncross_2021}. 
Except for KRb, each of these species has been predicted to be stable against chemical reactions in low-energy collisions of two molecules~\cite{Zuchowski_2010}. In spite of this, universal two-body loss, where almost every collision leads to the loss of both molecules, has been observed consistently in experiments~\cite{Takekoshi_2014, Guo_2018b, Ye_2018, Gregory_2019, Voges_2020}. 
We mean by ``universal'' that the probability of loss inside the short-range part of the interaction potential is unity. In this case, the collision process does not depend on the details of the short-range potential, and the loss rate is determined only by the long-range interactions that govern the flux of molecules reaching short range~\cite{Idziaszek_2010}.
 Strong loss has also been found with $^{40}$Ca$^{19}$F molecules in optical tweezers~\cite{Cheuk_2020}. Confining the molecules in very deep optical lattices~\cite{Chotia_2012, Moses_2015, Reichsoellner_2017} or using engineered repulsive interactions based on external dc or ac fields~\cite{Gorshkov_2008, Quemener_2010, Gonzalez-Martinez_2017, Lassabliere_2018, Karman_2018a, Xie_2020, Valtolina_2020, Matsuda_2020, Anderegg_2021, Li_2021} allow one to mitigate such losses by suppressing close-range collisions. Elastic collisions for the purpose of evaporative cooling can alternatively be established via collisions with atoms~\cite{Akerman_2017, Son_2020, Jurgilas_2021}.

However, the collisional loss still represents a major obstacle in the development of the field, as it makes evaporative cooling difficult and limits the lifetime of dense molecule samples. It has been suggested that the two-body loss may be caused by sticky collisions, where two molecules form a long-lived, intermediate four-body complex, which is subsequently lost from the trap. The first hypothesis was that these complexes could then undergo collisions with molecules~\cite{Mayle_2012, Mayle_2013, Croft_2014}. However, recent calculations predicted that the complex sticking time is too short for this to explain observations~\cite{Christianen_2019a, Christianen_2019b}. Instead, it was proposed that complexes could be excited by photons from the dipole trap, causing them to be transferred to states which do not decay back into ground-state molecules. Recent experiments with $^{40}$K$^{87}$Rb and $^{87}$Rb$^{133}$Cs, in optical dipole traps that were chopped on and off to be temporally dark, have provided evidence for this hypothesis~\cite{Liu_2020, Gregory_2020}. 

In this work, we characterize the two-body collisions of fermionic, chemically stable $^{23}$Na$^{40}$K molecules under highly controlled conditions of varying light intensity, temperature and electric field. Specifically, we realize trapping conditions of permanently very low light intensity in a blue-detuned optical box trap and compare the results to those from standard, red-detuned optical dipole traps. Under such low-intensity conditions, the lifetime of molecules should be significantly increased, as most sticky complexes are not excited by photon scattering and can therefore decay back into diatomic molecules. Surprisingly, and in stark contrast to the results found with other bialkali species, we find no dependence of the collisional behavior on the trapping light intensity. Our findings give a joint lower bound on the lifetime and photoexcitation rate of sticky complexes which disagrees by at least two orders of magnitude with state-of-the-art theoretical predictions.

\section{Theory of sticky collisions}
In a collision between two bialkali dimers, the strong chemical interaction energies (1000s of cm$^{-1}$) enable the colliding molecules to access a large number of rovibrationally excited states. With the molecules initially in their absolute ground state, the emergent four-body collision complex can only dissociate when both molecules have returned to the ground state. This leads to a complex sticking time $\tau_\mathrm{stick}$ which is much longer than is common in higher-temperature collisions~\cite{Hu_2019}. If the dynamics of the collision is chaotic, the sticking time can be estimated from the density of states of the complex $\rho_s$ via Rice-Ramsperger-Kassel-Marcus (RRKM) theory~\cite{Mayle_2012, Mayle_2013}:
\begin{equation}
\tau_{\mathrm{stick}} \approx \tau_{\mathrm{RRKM}}=\frac{2 \pi \hbar \rho_s}{N_s},
\end{equation}
where $N_s$ is the number of energetically available outgoing quantum states, with $N_s = 1$ for nonreactive molecules. In Ref.\ \cite{Christianen_2019a} the sticking time for $^{23}$Na$^{40}$K-$^{23}$Na$^{40}$K collisions was calculated using this formalism to be $\tau_{\mathrm{RRKM}} = \SI{18}{\mu s}$. This sticking time can be extended in the presence of external fields or rotation-hyperfine couplings which break conservation of motional angular momentum. 

\begin{figure}
\centering
\includegraphics{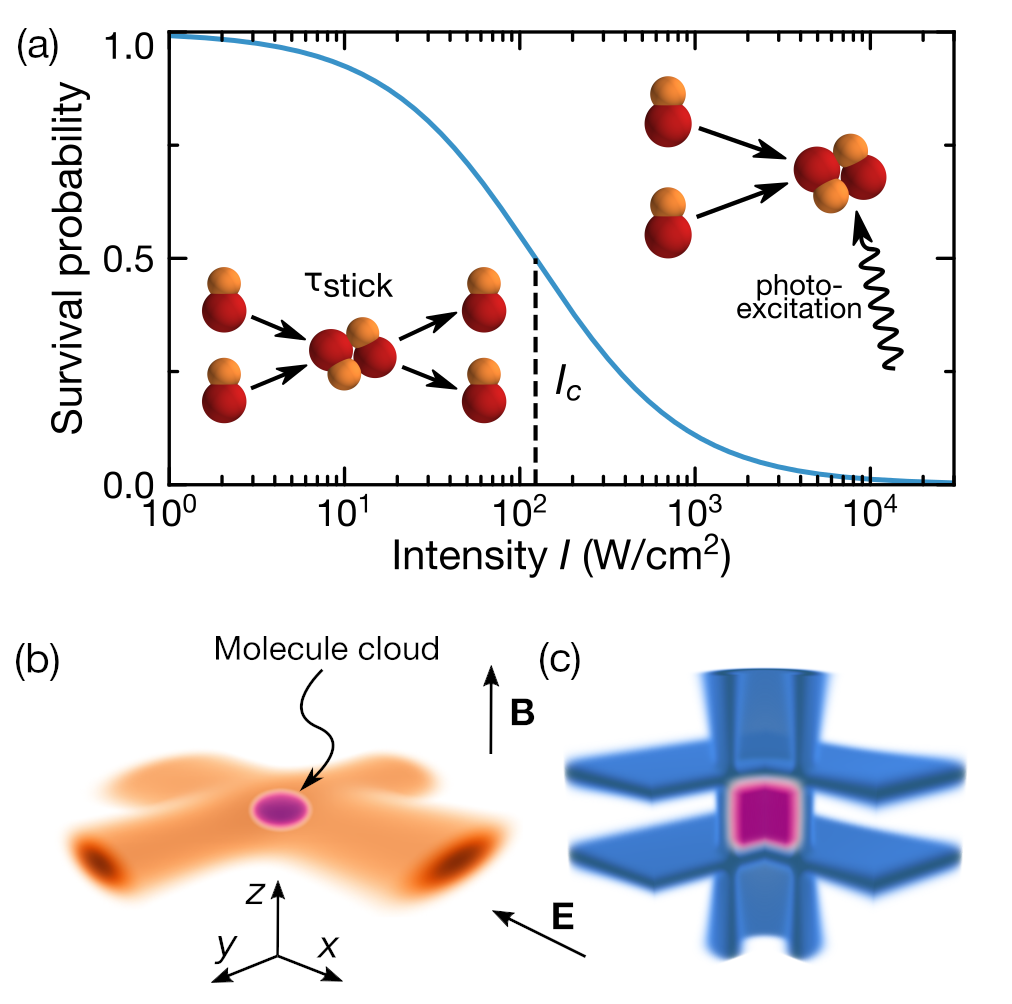}
\caption{Overview of sticky collisions and the experimental setup. (a)~Probability for two molecules to survive a collision at a given intensity of 1064-nm light according to Ref.~\cite{Christianen_2019a}. An intermediate sticky complex is formed in every collision, but at high intensity, almost every complex is lost via photoexcitation. Only at low intensity, the complex can decay back into ground-state molecules after a mean sticking time $\tau_{\mathrm{stick}}$. At the critical intensity $I_c$, the survival probability is 50\%. (b)~Red-detuned crossed trap consisting of two elliptical beams of wavelength \SI{1064}{nm} and \SI{1550}{nm}. (c) Blue-detuned box trap at \SI{866}{nm} with vertical cylinder beam and two horizontal elliptical beams. One quadrant is cut out for visibility. The arrows that indicate magnetic field and electric field are valid for both (b) and (c). To compensate gravity in the box trap, we use an electric-field gradient along the $z$ direction.}
\label{fig-overview}
\end{figure}

A molecular sticking of $\tau_{\mathrm{stick}}=\SI{18}{\mu s}$ by itself is not sufficient to account for the observed  two-body loss, since the complexes have enough time to dissociate into diatomic molecules before being lost from the trap. However, it was predicted in Ref.\ \cite{Christianen_2019b} that collision complexes can be electronically excited by photons from the trapping laser. After excitation, the complex can spontaneously decay into a large number of different quantum states, making a dissociation back into ground-state molecules unlikely. The calculated excitation rate for typical laser intensities in a red-detuned dipole trap is at least an order of magnitude faster than the dissociation rate of the complexes. It has therefore been the accepted explanation for the experimentally observed two-body loss of molecules (see Fig.~\ref{fig-overview}).

\section{Experiments in the red-detuned dipole trap}
\subsection{Molecule preparation and trapping}

We typically create $2.5\times10^4$ $^{23}$Na$^{40}$K ground-state molecules at temperatures between \SI{300}{nK} and \SI{1}{\mu K} from an ultracold atomic mixture that contains about $2 \times 10^5$ potassium and $1 \times 10^5$ sodium atoms. The mixture is initially prepared in a crossed-beam optical dipole trap, which consists of two orthogonal laser beams with wavelengths of \SI{1550}{nm} and \SI{1064}{nm}, as illustrated in Fig.~\ref{fig-overview}(b). 
The Na and K atoms are in the states $|F, m_F\rangle = |1, 1\rangle$ and $|9/2, -9/2\rangle$, respectively, where $F$ is the total atomic angular momentum and $m_F$ is its projection onto the magnetic field axis. We then ramp the external magnetic field over the interspecies Feshbach resonance at \SI{78.3}{G}~\cite{Park_2012} in order to form weakly bound molecules in the least-bound vibrational state of an electronic manifold with mainly $a^3\Sigma^+$ character. 
Using a stimulated Raman adiabatic passage (STIRAP) pathway similar to the one described in Refs.\ \cite{Park_2015, Park_2015a, Liu_2019b}, we transfer typically 80\% of these molecules into the rovibrational and hyperfine ground state $|J, m_J, m_{I,\rm{Na}}, m_{I,\rm{K}}\rangle = |0, 0, 3/2, -4\rangle$ of the electronic $X^1\Sigma^+$ manifold. Here, $J$ denotes the total molecular angular momentum excluding nuclear spin \cite{footnote_1}, $m_J$ is its projection onto the quantization axis, and $m_{I,\rm{Na}}$ ($m_{I,\rm{K}}$) is the nuclear spin projection of Na (K). In order to study the collisional properties of the molecules, we keep them confined in the crossed dipole trap at trap frequencies of $(\omega_x, \omega_y, \omega_z) = 2\pi \times (89(2), 57(1), 205(3)) \, \mathrm{Hz}$. The trap depth is $k_B \times \SI{6}{\mu K}$. Unless otherwise stated, the STIRAP and holding of ground-state molecules are performed at \SI{72.4}{G}, where the STIRAP two-photon detuning is insensitive to small changes in the magnetic field. At the end of the experimental run, we reverse the STIRAP and subsequently perform absorption imaging.

\subsection{Data analysis}
With the molecule density distribution expected in a harmonic dipole trap, the remaining molecule number after a given hold time, $N(t)$, is described by the differential equation
\begin{equation}
\frac{dN}{dt} = -\frac{\beta N^2}{\nu_0 T^{3/2}}.
\label{eq-loss-differential-equation}
\end{equation}
Here $\nu_0T^{3/2}$ is the generalized trap volume given by $\nu_0 = (4\pi k_B / m\bar{\omega}^2)^{3/2}$, with $\bar{\omega} = (\omega_x\omega_y\omega_z)^{1/3}$ the geometric mean of the trap frequencies and $m$ the molecule mass. The two-body loss coefficient is denoted by $\beta$. During the holding time, previous experiments have observed a linear increase in temperature over time such that $T(t) = T_0 + qt$~\cite{Ospelkaus_2010a, Ni_2010a}. This heating is caused by the lowest-energy molecules being lost predominantly, as they are located in the trap center where the density is highest. With such heating present, the solution of Eq.\ (\ref{eq-loss-differential-equation}) is
\begin{equation}
N(t) = \frac{N_0}{1 + \frac{2\beta N_0}{q \nu_0} \left(\frac{1}{\sqrt{T_0}} - \frac{1}{\sqrt{T_0 + qt}}\right)},
\label{eq-loss-fit}
\end{equation}
where $N_0$ is the initial molecule number and $\beta$ is assumed to be constant. This is a reasonable assumption because the largest observed temperature change is below 25\% over the hold times used in the experiment. Importantly, the model assumes thermal equilibrium of the molecule sample at all times, and, even though thermalization is impossible without elastic collisions, it still describes our observations remarkably well. 

\subsection{Temperature dependence of collisions}
\begin{figure}
\centering
\includegraphics{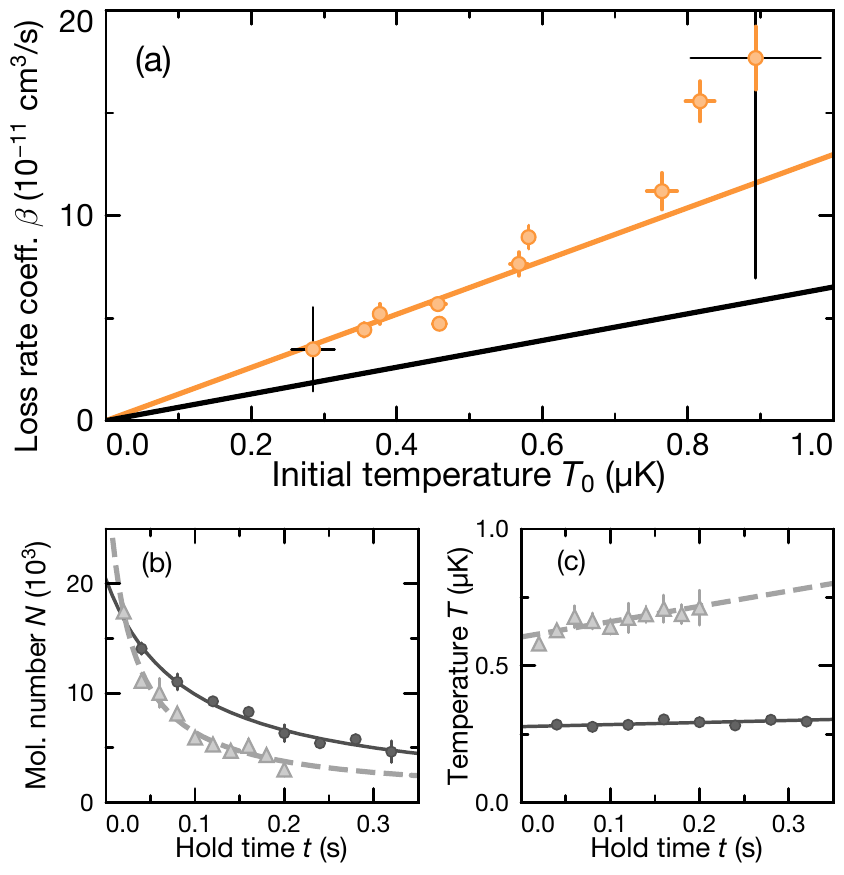}
\caption{Temperature dependence of two-body loss coefficient $\beta$. (a)~Measured $\beta$ versus initial temperature of the molecule sample in the crossed dipole trap. The orange line is a linear fit to the data resulting in a slope of $\beta/T = \SI{13.0(8)e-11}{cm^3/\mu K s}$, compared to $\beta/T = \SI{6.52e-11}{cm^3/\mu K s}$ obtained from the MQDT calculation without fit parameters (black line). The orange error bars represent the 1$\sigma$ uncertainty of the fit, and the black error bars additionally contain the systematic uncertainty. (b)~Sample data for molecule number versus hold time for two different initial temperatures (triangles and circles) and two-body loss fits (solid lines) according to Eq.\ (\ref{eq-loss-fit}). (c)~Sample data for sample temperature versus hold time of the same two data sets shows a linear dependence of temperature on hold time. For (b) and (c), error bars correspond to the standard error of the mean of three repetitions.}
\label{fig-loss-rate-vs-temperature}
\end{figure}
For identical fermionic molecules, we expect $p$-wave two-body collisions to give rise to the dominant loss mechanism. As previously observed with fermionic $^{40}$K$^{87}$Rb, the two-body loss coefficient $\beta$ should follow $\beta\propto T$ at temperatures much lower than the height of the $p$-wave barrier~\cite{Ni_2010a}. We experimentally confirm this by holding molecules in the crossed dipole trap for a hold time $t$ at different initial temperatures $T_0$ and comparing the remaining molecule number to our model (see Fig.~\ref{fig-loss-rate-vs-temperature}).

From a multichannel quantum defect theory (MQDT) calculation using the universal condition, Ref.\ \cite{Ospelkaus_2010a} found $\beta = (11.48 \, \bar{a})^3(k_BT/h)$ with the characteristic van-der-Waals length $\bar{a} = \SI{12.7}{\nano\meter}$~\cite{Lepers_2013}. This result is approximately two times smaller than our experimentally-found $\beta$. One possible explanation for this discrepancy could lie in our density calibration, which is subject to a systematic uncertainty estimated to be up to 50\% because we determine density indirectly from the measured molecule numbers, trap frequencies, and temperatures. It is also possible that the collisions of nonreactive molecules are nonuniversal, as it was found for $^{23}$Na$^{87}$Rb and $^{87}$Rb$^{133}$Cs. In this case, interference between ingoing and reflected parts of the molecular wavefunction in a collision can modify the loss rate, so that $\beta/T$ deviates from the universal MQDT expression~\cite{Ye_2018, Bai_2019, Gregory_2019}.
 
\subsection{Dipolar collisions}
\begin{figure}
\centering
\includegraphics{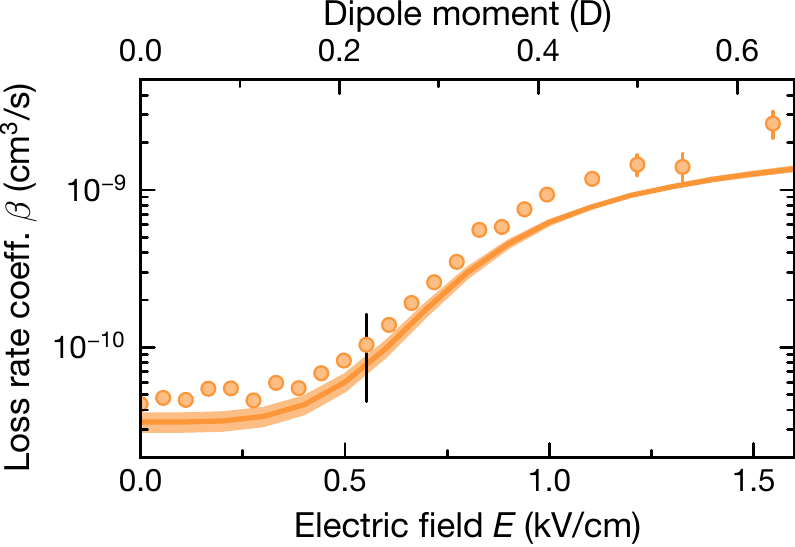}
\caption{Loss rate coefficient $\beta$ depending on external electric field in the red-detuned crossed dipole trap at temperatures between 480 and $\SI{620}{nK}$. The solid line shows a parameter-free quantum close-coupling calculation at $\SI{550}{nK}$. The shaded area around the theory curve takes the highest and lowest observed temperatures into account. The orange error bars correspond to the 1$\sigma$ uncertainty of the fits. The black error bar additionally contains the systematic uncertainty.}
\label{fig-loss-rate-vs-efield-ct}
\end{figure}
Polarizing the molecules by an external dc electric field induces a dipole-dipole interaction between them which strongly modifies their collision behavior: side-by-side collisions are suppressed by repulsion; at the same time, the attractive interaction in head-to-tail orientation weakens the $p$-wave barrier and increases the collision rate. 
Using four rod electrodes situated inside the ultrahigh vacuum chamber, we applied dc electric fields to the molecule samples and determined loss rate coefficients as described previously. We again assumed $\beta$ to be constant in each fit, because at high electric fields, $\beta$ depends only weakly on $T$ due to the significantly reduced $p$-wave barrier. Hence, the effect of the temperature-dependence of $\beta$ is weak, even though the maximum temperature can increase by up to 50\% during our hold times.
Fig.\ \ref{fig-loss-rate-vs-efield-ct} shows our observed dependence compared to a numerical coupled-channel calculation (see Appendix \ref{appendix-cc-calculation}). We find overall agreement between theory and experiment up to a constant factor $\simeq 1.4$, possibly also resulting from uncertainties in our density calibration.

\section{Experiments in the box trap}
\subsection{Trap setup}
Previous experiments have used the method of chopped, temporally dark dipole traps to investigate the intensity-dependence of loss~\cite{Liu_2020, Gregory_2020}. This method is fundamentally limited to relatively short dark times due to parametric heating which becomes strong once the chopping frequency is on the same order as the harmonic trap frequency. 
In contrast, by using a repulsive dipole force, molecules can be trapped in regions of continuously low light intensity. Though trapping of atoms in  box potentials has been demonstrated previously~\cite{Gaunt_2013, Mukherjee_2017}, achieving a repulsive optical dipole force for molecules is nontrivial. This is because it is difficult to find laser wavelengths which are sufficiently far blue-detuned from any one transition without being too close to higher-lying transitions.  
In a recent study, we have characterized the nominally forbidden transition $|X^1\Sigma^+, v=0, J=0\rangle \leftrightarrow |b^3\Pi_0, v=0, J=1\rangle$ at a wavelength of \SI{866.1428(3)}{\nano\meter}, which exhibits a linewidth much smaller than its separation to the next higher-lying transition~\cite{Bause_2020}. Working \SI{300}{\mega\hertz} blue-detuned from this transition thus allows us to apply a repulsive dipole force while keeping photon-scattering rates small. 

The employed light is generated by a Ti:sapphire laser locked to a Fizeau-interferometer-based wavemeter. A hollow ring-shaped beam with a radius of $\SI{60}{\mu m}$ \cite{Manek_1998}, and two strongly elliptical beams at a distance of $\SI{75}{\mu m}$ together form the cylindrical box potential, as shown in Fig.~\ref{fig-overview}(c). Masks in the beam paths, which block light in the low-intensity regions, are imaged onto the molecules. The average light intensity experienced by a molecule in the trap can thereby be reduced to $\SI{0.70(25)}{W/cm^2}$, 6000 times lower than is typical for our crossed dipole trap. The method by which we determined this value is described in Appendix~\ref{appendix-residual-intensity}. At this intensity, the photon-scattering rate is \SI{0.12(4)}{Hz}, and we achieve a trap depth of $k_B \times \SI{3}{\mu K}$. Hence, the box trap is an ideal tool for studying the intensity-dependence of the complex lifetime. Creating a homogeneous potential for the molecules additionally requires compensation of gravity, which can be achieved with an  electric offset field that linearly increases in magnitude along the $z$ direction. The electrodes were used to create such an electric-field distribution by applying a superimposed dipole and quadrupole voltage configuration. The box trap was loaded by first ramping up the electric field for levitation and turning on the box trap with the molecules still trapped in the crossed dipole trap. Afterwards, the depth of the crossed dipole trap was adiabatically reduced to zero to transfer the molecules with high efficiency into the box trap, where they ended up with a temperature of typically \SI{140}{nK}. Because the box trap relies on near-resonant trapping of a specific rovibrational state, residual atoms and molecules in other rovibrational states are not trapped, ensuring a pure sample.

\subsection{Data analysis}
The density distribution in the box trap is nearly homogeneous. Therefore, in contrast to the situation in the harmonic crossed dipole trap, there should be no heating effect caused by colder molecules colliding more frequently. Indeed, we did not observe temperature changes during the hold time. However, as the box trap is operated with near-detuned light, the one-body loss due to photon scattering must be taken into account, such that Eq.\ (\ref{eq-loss-differential-equation}) becomes
\begin{equation}
\frac{dN}{dt} = -\frac{\beta N^2}{V} - \Gamma_{\mathrm{sc}}N,
\label{eq-loss-differential-equation-box}
\end{equation}
where $V$ is the trap volume, which is numerically determined from the known intensity profile of the box trap, and $\Gamma_{\rm{sc}}$ is the measured photon-scattering rate, which is much smaller than the initial two-body loss rates (see Appendix \ref{appendix-residual-intensity}). The solution in this case is
\begin{equation}
\label{eq-loss-fit-box}
N(t) = \frac{N_0 e^{-\Gamma_{\mathrm{sc}}t}}{1 + \frac{N_0 \beta }{ V \Gamma_{\mathrm{sc}}} (1 - e^{-\Gamma_{\mathrm{sc}} t})}.
\end{equation}
In the box trap, the distribution of molecules can be determined more precisely than in the crossed trap, as it depends only weakly on temperature. However, there is increased systematic uncertainty in the molecule number. Since the STIRAP beam size is comparable to the box-trap size, the conversion efficiency of ground-state molecules back to the Feshbach-molecule state for detection is reduced compared to the crossed trap. Because Feshbach molecules are not trapped in the box trap, we cannot distinguish between reduced STIRAP efficiency and imperfect loading. A further systematic effect is caused by the uncertainty of  $\Gamma_{\mathrm{sc}}$.

\subsection{Light-intensity dependence of collisional loss} 
\begin{figure}
\centering
\includegraphics{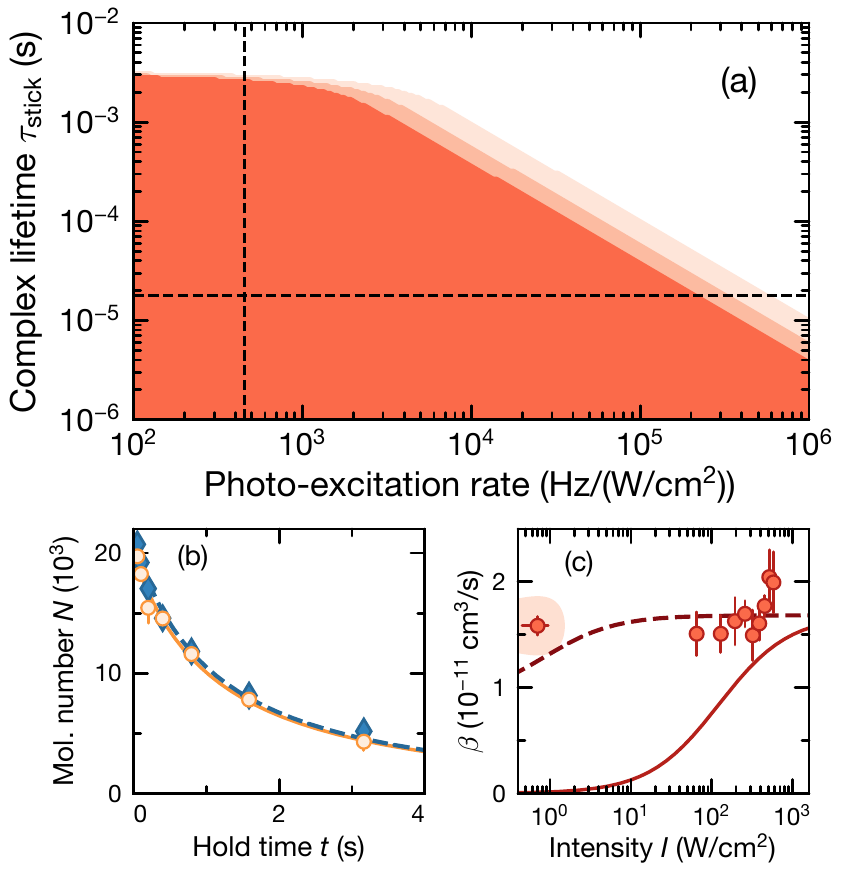}
\caption{Intensity dependence of loss in the levitated box trap. (a) Parameter space excluded by our data. The three shaded areas indicate regimes that are excluded with $1\sigma$, $2\sigma$, and $3\sigma$ confidence, with darker color corresponding to higher confidence. The dashed lines indicate the predictions from Ref.~\cite{Christianen_2019b}. Our method becomes insensitive at complex lifetimes above \SI{3}{ms} because here, all complexes fall out of the trap regardless of photoexcitation probability. (b) Comparison of loss with kill-beam off (blue diamonds) and on at an intensity of $I=\SI{276}{W/cm^2}$ (orange circles). The dashed (solid) lines are fits of Eq.\ (\ref{eq-loss-differential-equation-box}) to determine $\beta$ with the kill-beam off (on). The error bars represent the standard error of the mean of three repetitions. (c) Dependence of $\beta$ on $I$. For comparison, the solid line shows the expected value of $\beta$, assuming $\tau_{\mathrm{stick}} = \tau_{\mathrm{RRKM}}$, while the dashed line assumes $\tau_{\mathrm{stick}} = \SI{2.6}{ms}$. For both curves it is assumed that complexes are not trapped by the box trap. The shaded area indicates the 3$\sigma$ confidence region of the lowest-intensity data point. The error bars indicate the 1$\sigma$ uncertainty of the fit.}
\label{fig-loss-levitated-box}
\end{figure}

With the low intensity experienced by molecules in the levitated box trap, a suppression of the loss rate to a value significantly lower than the universal loss rate would be expected (see Fig.~\ref{fig-overview}). To be able to directly compare low- and high-intensity trapping conditions, we used a large-diameter, non-modulated, 1064-nm laser beam with high intensity $I$ (subsequently called the ``kill-beam''). We measured loss rate coefficients both with the kill-beam blocked and with values of kill-beam intensity up to $\SI{300}{W/cm^2}$ at an offset electric field of $E=\SI{411}{V/cm}$. We did not observe any difference in the loss behavior (see Fig.\ \ref{fig-loss-levitated-box}). The kill-beam radius is $\SI{350}{\mu\meter}$, sufficiently large to exclude density changes due to an additional dipole trap effect. We then compared our results to a simple model in which complexes can only be lost either by photoexcitation with the theoretically predicted rate or by leaving the trap (see Appendix~\ref{appendix-complex-loss}). With this model we find that the smallest sticking time that is consistent with our lowest-intensity data within $3\sigma$ is $\tau_{\mathrm{stick}} = \SI{2.6}{ms}$, 140 times larger than $\tau_{\mathrm{RRKM}}$, assuming that the model predicts the photoexcitation rate accurately. Fig.~\ref{fig-loss-levitated-box}(a) shows the experimentally excluded parameter regime for a variable photoexcitation rate.

In Ref.\ \cite{Christianen_2019a} it was shown that the sticking time can be enhanced by orders of magnitude by external fields, because breaking the conservation of angular momentum leads to a large increase of the density of states. However, it is unclear how large electric or magnetic fields need to be for this effect to occur. To exclude a possible effect of the electric field on the complex lifetime, we repeated our measurements of the intensity-dependence of the loss rate without applying voltage to the electrodes, which results in electric fields of \SI{1}{V/cm} or below. Our results again show no difference between dark and bright trapping conditions within our measurement precision, see Appendix \ref{appendix-non-levitated-box}. In this case the molecules can not be levitated, causing complexes to be formed mostly at the bottom of the trap, such that they leave the trap at a faster rate. Hence our data only allows us to conclude that $\tau_{\mathrm{stick}} > \SI{1.4}{ms}$ at zero electric field. 
We consider it unlikely that the small background electric fields in the nonlevitated box trap would be sufficient to cause a drastic increase of the sticking time, since this was also not observed in previous experiments~\cite{Gregory_2020}. We consider it equally unlikely that the presence of the magnetic fields causes the sticking time enhancement, as further experimental checks also revealed no dependence of loss rates on the magnitude of the magnetic background field for values between 10 and \SI{100}{G}.

\subsection{Dipolar collisions}
\begin{figure}
\centering
\includegraphics{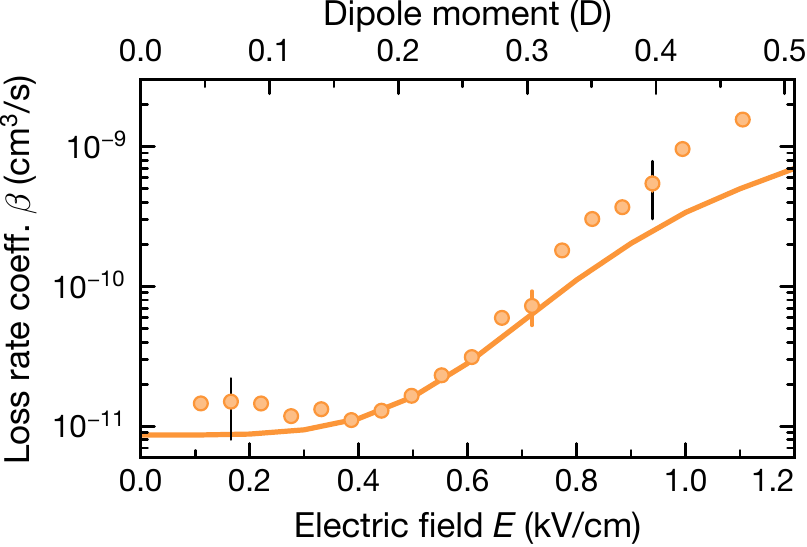}
\caption{Loss rate coefficient $\beta$ depending on external electric field $E$ in the box trap with electric levitation at $T=\SI{140}{nK}$. The solid line is a parameter-free quantum close-coupling calculation. The orange error bars represent the 1$\sigma$ uncertainty of the fit. The black error bars also contain the systematic uncertainty from density calibration and photon-scattering rate.}
\label{fig-loss-rate-vs-efield_box}
\end{figure}

We then repeated the measurements of the dependence of $\beta$ on the dc electric field in the box trap with levitation (see Fig.\ \ref{fig-loss-rate-vs-efield_box}). As previously observed in the crossed dipole trap, the loss rates are consistent with near-universal loss over the entire investigated range of electric fields. The laser frequency of the box-trap light was not changed for varying $E$. 
This affects the detuning from the molecular transition, as the transition frequency is increased due to a dc Stark shift. However, at $E < \SI{1.2}{kV/cm}$, the measured $\beta$ was unchanged for a detuning doubled to \SI{600}{MHz} with unchanged trap power, proving that an increased photon-scattering rate is not responsible for the increased loss seen at high electric fields. 

\subsection{Enhanced sticking time due to $p$-wave barrier}
The theory in Ref.\ \cite{Christianen_2019a} only describes the behavior of the complex at short range. Here, we propose that the long-range part of the potential, and in particular, the centrifugal barrier, can have a strong effect on the effective sticking time. The idea is simple: just like molecules need to tunnel through the centrifugal barrier to enter the short-range part of the potential, they also need to tunnel through the centrifugal barrier to leave the short-range part of the potential again~\cite{Gao_2010, Wang_2012}. This means that the molecules do not just form the sticky short-range complex once, but multiple times, depending on the collision energy. The effective sticking time observed in the experiment then becomes $\tau_\mathrm{stick} \approx \tau_\mathrm{RRKM} \mathcal{N}$ with the recollision number $\mathcal{N}$. The number of recollisions is estimated to be $\mathcal{N}_{s\text{-wave}}=3$ for the $s$-wave case and $\mathcal{N}_{p\text{-wave}}=270$ for the $p$-wave case (see Appendix\ \ref{appendix_recollisions}). Using this value, the resulting prediction is consistent with the data shown in Figs.\ \ref{fig-loss-levitated-box} and \ref{fig-loss-non-levitated-box}. This could also account for the difference with the $^{87}$Rb$^{133}$Cs and $^{40}$K$^{87}$Rb experiments, where this recollision effect plays a much smaller role, because $^{87}$Rb$^{133}$Cs is bosonic and barrierless and for $^{40}$K$^{87}$Rb most of the outgoing channels are reactive. 
\begin{figure}
\centering
\includegraphics{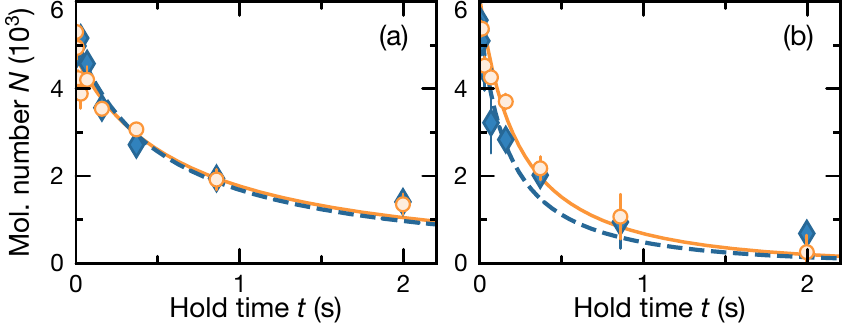}
\caption{Comparison between low- and high-intensity trapping of an incoherent mixture of the states $|0, 0, 3/2, -4\rangle$ and $|0, 0, 3/2, -3\rangle$ in the box trap. (a)~Remaining molecule number versus hold time with electric levitation. Dark diamonds are taken without the kill-beam, bright circles are with kill-beam intensity $I=\SI{204}{W/cm^2}$. The solid (dashed) line is a fit of Eq.\ (\ref{eq-loss-fit-box}) to the data with kill-beam on (off). Error bars represent the standard error of the mean of three repetitions. (b)~Like (a), but at zero electric field without levitation.}
\label{fig-mixture}
\end{figure}

To experimentally test the recollision hypothesis, we created an incoherent mixture of molecules in the two lowest hyperfine states, $|0, 0, 3/2, -4\rangle$ and $|0, 0, 3/2, -3\rangle$~\cite{Ospelkaus_2010a}. Since the fermionic molecules are then no longer identical, collisions predominantly have $s$-wave character, leading to a much smaller expected effect of recollisions. In addition, $\tau_{\mathrm{RRKM}}$ is three times smaller for the $s$-wave case. 

The mixture was created from a pure ground-state sample with two microwave pulses. The first pulse creates a coherent superposition $|0, 0, 3/2, -4\rangle + |1, 1, 3/2, -4\rangle$, which is held until the dipole trap light has led to complete decoherence of the superposition. The second pulse then transfers all population from $|1, 1, 3/2, -4\rangle$ to $|0, 0, 3/2, -3\rangle$. In order to not create additional unwanted decoherence during the microwave pulses, all dipole traps were switched off during the pulse duration. 
We then repeated the loss measurements in the box trap, again with and without the kill-beam present and both with electric levitation and at zero electric field (see Fig.\ \ref{fig-mixture}). As expected, the collision rate is significantly increased because of the $s$-wave contribution. However, we do not see any evidence for a shorter sticking time of $s$-wave complexes. The smallest values of $\tau_{\mathrm{stick}}$ which are consistent with our observations within 3$\sigma$ are \SI{2.3}{ms} for the levitated case and $\SI{133}{\mu s}$ for the zero-field case. Under the conditions of this experiment, about 80\% of complexes should be formed in $s$-wave collisions. According to the RRKM theory with recollisions, the lifetime of these complexes is expected to be $\SI{18}{\mu s}$ because the effect of recollisions approximately cancels out the shorter $\tau_{\mathrm{RRKM}}$ of $s$-wave complexes. Hence, even though the recollision hypothesis may hold, it can not fully explain our experimental results.

\section{Possible explanations}
Our results can in principle be explained in two qualitatively different ways: either photoexcitation of the complexes is indeed the main source of the molecule loss, but the effective complex loss is at least two orders of magnitude larger than expected; or there is an additional loss mechanism that is dominant at low light intensities and that does not require laser excitation. The former could happen if either the photoexcitation rate of the complex, or the sticking time, or a combination of both, was underestimated in Ref.\ \cite{Christianen_2019b}. The predicted photoexcitation rates have large uncertainties due to the difficulty of calculating transition-dipole-moment surfaces, however, we do not expect the uncertainties in these rates to be large enough to explain our results. 

On the other hand, $\tau_{\mathrm{RRKM}}$ could be larger by multiple orders of magnitude if hyperfine spin flips are possible within the complex. These, in principle, could be caused by coupling between the electric-field gradient and the nuclear quadrupole moment~\cite{Aldegunde_2017} and would increase the number of accessible states for the complex significantly. The time scale on which these spin flips would take place is unknown. However, because they have not been observed in the cases of $^{40}$K$^{87}$Rb or $^{87}$Rb$^{133}$Cs~\cite{Hu_2020, Gregory_2020}, we consider this unlikely to be an important factor.

The number of accessible states inside the sticky complex can also be increased significantly in presence of strong external magnetic or electric fields \cite{Christianen_2019b}. Again, it is so far unclear what the critical value for these field strengths would be, but we did not find any change in collision rates, even at magnetic fields of \SI{10}{G} and electric fields of less than \SI{1}{V/cm}. A recent experiment with $^{40}$K$^{87}$Rb also did not detect any influence of small fields on collisions~\cite{Liu_2021}.

To exclude the possibility of a problem with the box trap itself, we also performed experiments in a chopped dipole trap. The results are described in Appendix~\ref{appendix-chopped-trap} and confirm the lack of intensity dependence of the loss.

Loss mechanisms other than laser excitation that could explain the observations include collisions of the complex with a third ground-state molecule, escape of the complexes from the trap, and relaxation of the complex by spontaneous photoemission. However, we estimate the rates of these processes to be orders of magnitude slower than the complex dissociation rate (without recollisions), making it unlikely that they can explain our results. In the presence of recollisions for fermionic molecules or an otherwise increased sticking time, some of these processes may play a role. Furthermore, it is possible that there is a loss mechanism unaccounted for, which is dominant under our experimental conditions.

\section{Discussion and Conclusions}
We have demonstrated that $^{23}$Na$^{40}$K molecules exhibit near-universal two-body loss even under conditions of very low light intensity. This was enabled by loading molecules into an optical box trap with extremely low residual light intensity. Though in this study, the purpose of the box trap was only to enable low-intensity trapping, we note that it also provides a homogeneous potential for molecules. This may allow future studies of e.g.\ the Fermi surface of strongly dipolar gases without harmonic confinement. 

Assuming that complexes are lost only by photon scattering or leaving the trap, we determined a lower bound for the average sticking time of $\tau_{\mathrm{stick}} = \SI{2.6}{ms}$ and $\tau_{\mathrm{stick}} = \SI{1.4}{ms}$, respectively, in the levitated and in the electric-field-free case. These values are two orders of magnitude larger than predicted. Our results also show that the loss can not be reduced below the universal limit using different magnetic and electric background fields, different molecular hyperfine states, or with dominant $s$-wave instead of $p$-wave collisions. We conclude that the recent theoretical model of intensity-dependent loss of sticky collision complexes in Refs.~\cite{Christianen_2019a, Christianen_2019b} misses crucial ingredients in the description of $^{23}$Na$^{40}$K-$^{23}$Na$^{40}$K collisions. The hypothesis that recollisions increase the effective lifetime of four-body complexes could resolve this discrepancy for the case of identical fermions, but fails to explain the results obtained with hyperfine mixtures. Indeed, even for a mixture held without any electric field, we still find $\tau_{\mathrm{stick}} > \SI{133}{\mu s}$, seven times larger than expected after taking recollisions into account.

In parallel with our work, similar results were found with $^{23}$Na$^{39}$K and $^{23}$Na$^{87}$Rb in temporally dark traps~\cite{Gersema_2021}. As both molecules are bosonic, recollisions are insufficient to explain these observations. In combination with our results, the new findings are particularly relevant because sticky collisions---and thus collisional loss of ultracold molecules---were believed to be understood since recent experiments on $^{87}$Rb$^{133}$Cs and $^{40}$K$^{87}$Rb had shown good agreement with theory predictions~\cite{Gregory_2020, Liu_2020}. This raises the question of whether there is a yet unknown property that differentiates the molecular species. Further investigation of this conundrum will be crucial for the advancement of ultracold molecular research. It is likely that the creation of high-density molecular gases in three dimensions, including the long-sought Bose--Einstein condensate of dipolar molecules, will only be possible with a proper understanding of collisions. Such an understanding may also lead to simpler and more efficient evaporative cooling schemes, enabling the creation of deeply degenerate samples.

\begin{acknowledgments}
We thank Gerrit Groenenboom for insightful discussions and Yang Cui for help with producing custom optical masks. The MPQ team gratefully acknowledges support from the Max Planck Society, the European Union (PASQuanS Grant No. 817482) and the Deutsche Forschungsgemeinschaft, grants No. EXC-2111--390814868 and FOR 2247. G.Q. acknowledges funding from the FEW2MANY-SHIELD Project No. ANR-17-CE30-0015 from Agence Nationale de la Recherche.

R.B., A.S. and M.D. conducted the experiments, R.T., X.-Y.L. and R.B. set up the box-trap apparatus, R.B., A.S. and X.-Y.C. analyzed the experimental data. G.Q. conducted the simulations on the electric-field dependence of collision rates, T.K. and A.C. performed the calculations on complex decay. I.B. and X.-Y.L. supervised the experiment. All authors contributed to the preparation of the manuscript.
\end{acknowledgments}

\appendix

\section{Quantum close-coupling calculation}
\label{appendix-cc-calculation}
The quantum close-coupling calculation is performed using a time-independent quantum formalism based on Jacobi coordinates. It includes the internal rotational structure of the molecules and is based on a partial-wave decomposition of the total wavefunction~\cite{Quemener_2018}. The formalism also includes an electric field which mixes different parities of the rotational states~\cite{Wang_2015}. At short range, a boundary condition is introduced for the wavefunction so that when the two molecules meet, they are considered lost with unit probability. As previous experimental results on fermionic $^{40}$K$^{87}$Rb and bosonic $^{23}$Na$^{87}$Rb in electric fields reported very good agreement with theoretical predictions using this universal condition~\cite{Ni_2010a, Guo_2018b}, we also adopt this assumption for the present study. A log-derivative type of propagation is used to solve the coupled Schr\"{o}dinger equations. At long range, usual asymptotic boundary conditions are applied, and one can extract the scattering matrix for a given collision energy and electric field. From this matrix, one can obtain the loss cross section for the initial state prepared in the experiment. To get the loss rate coefficient as a function of the temperature, the loss cross sections are multiplied by their relative velocities related to their collision energies. Then this product is averaged over a Maxwell-Boltzmann distribution of the velocities for the given temperature.

\section{Residual intensity in the box trap}
\label{appendix-residual-intensity}
\begin{figure}
\centering
\includegraphics{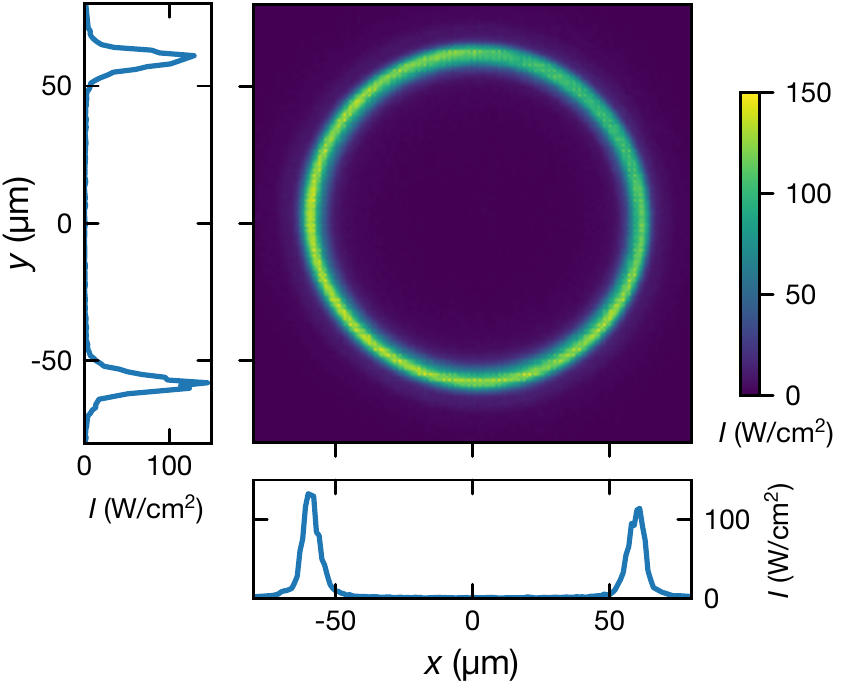}
\caption{Shape of the ring beam. Cuts through the beam profile through the center of the ring are shown on the left and bottom. Note that, due to scattered light and imaging noise, the residual intensity inside the trap is overestimated.}
\label{fig-ring-beam}
\end{figure}

\begin{figure}
\centering
\includegraphics{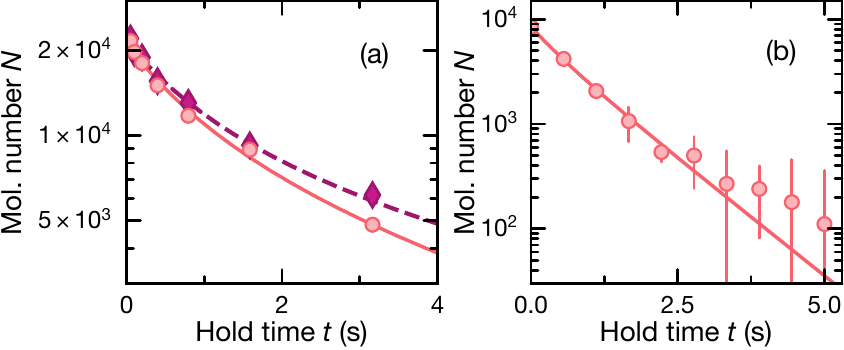}
\caption{Determination of one-body loss rate in the box trap. (a)~Molecule loss in the levitated case, with box-trap detuning \SI{300}{MHz} (\SI{600}{MHz}) shown as bright circles (dark diamonds). The solid and dashed lines show a simultaneous fit of Eq.\ (\ref{eq-loss-fit-box}) to both data sets. (b)~Nonlevitated case. The fit was used to determine $\Gamma_{\mathrm{sc}}$. Error bars represent the standard error of the mean of three repetitions.}
\label{fig-photon-scattering-rate}
\end{figure}

For illustrative purposes, an image of the ring beam is shown in Fig.~\ref{fig-ring-beam}. To determine the residual intensity in the box trap accurately and \textit{in situ}, both with and without levitation, the one-body loss rate was extracted from our data. In the levitated case, we compared datasets taken at two different detunings of the trap light, \SI{300}{MHz} and \SI{600}{MHz}. We then performed a simultaneous fit of Eq.\ (\ref{eq-loss-fit-box}) to both datasets, where we forced a quadratic dependence of $\Gamma_{\mathrm{sc}}$ on the detuning~\cite{Bause_2020}. This results in $\Gamma_{\mathrm{sc}} = \SI{0.12(4)}{Hz}$. For the nonlevitated box, we instead took data at long hold times, where the one-body loss to photon scattering outweighs the collisional loss and determined $\Gamma_{\mathrm{sc}} = \SI{1.0(3)}{Hz}$. The data and fits for both cases are shown in Fig.\ \ref{fig-photon-scattering-rate}.

These results were checked by microwave spectroscopy on the $|0, 0, 3/2, -4\rangle \leftrightarrow |1, 0, 3/2, -4\rangle$ transition. At a given intensity, the repulsive walls of the box trap cause a known light shift~\cite{Bause_2020}, which can be used to quantify the mean intensity experienced by the molecules by comparing with a case where the box trap is suddenly turned off before the microwave pulse. A histogram of the distribution of molecules versus light intensity was reconstructed from the microwave spectroscopy data. From this we determined $\Gamma_{\mathrm{sc}} = \SI{0.1}{Hz}$ for the levitated box trap and $\Gamma_{\mathrm{sc}} = \SI{1.0}{Hz}$ for the nonlevitated box trap, which is consistent within $1\sigma$ with the values from loss data. In Table~\ref{table-residual-intensity} we give an overview of the residual intensities in the box trap at \SI{866}{nm} as well as at all other wavelengths used in our experiment. 

\begin{table}
\caption{Residual intensity at the molecule position during the hold time in the box trap by wavelength.}
\begin{tabular}{l l l} 
\hline
   Wavelength (nm) & Intensity (W/cm$^2$)\\ 
   \hline
   1550 & $25 \times 10^{-6}$ \\
   1064 & $1.9 \times 10^{-3}$ \\
   866 (with levitation) & 0.70(25)\\
   866 (without levitation) & 6(2) \\
   805 & $1.3 \times 10^{-3}$ \\
   767 & $0.12 \times 10^{-3}$ \\
   589 & $50 \times 10^{-6}$ \\
   567 & $1.3 \times 10^{-3}$ \\
   \hline
\end{tabular}
\label{table-residual-intensity}
\end{table}

\section{Complex loss model}
\label{appendix-complex-loss}
We assume that there are two ways for complexes to be lost---either by photoexcitation or by leaving the trap. These processes are statistically independent, such that the probability of a sticky complex to survive and dissociate to diatomic molecules is $P_\mathrm{dis} = (1 - P_t) (1 - P_\mathrm{sc})$. Here, $P_t$ is the probability to leave the trap and $P_\mathrm{sc}$ is the photon-scattering probability. The latter can be estimated as follows:
\begin{equation}
P_\mathrm{sc} = \frac{1 / \tau_\mathrm{stick}}{1 / \tau_\mathrm{stick} + \gamma_l I}.
\end{equation}
Here, $\gamma_l$ denotes the complex photoexcitation rate, which was estimated in Ref.\ \cite{Christianen_2019b} to be $\gamma_l = \SI{452}{Hz/(W/cm^2)}$ for 1064-nm light. Though this rate may be higher at \SI{866}{nm}, the difference is likely small.

Estimating the loss due to complexes leaving the box trap requires knowledge of the distribution of their initial positions and momenta. These are found by assuming that the complexes are formed in thermal equilibrium with the molecules, with a homogeneous density inside the trap. As it is unlikely that the complexes exhibit the same near-resonant response to the trapping light as free molecules, it is a reasonable assumption that the box trap does not confine them. They can also not be electrically levitated because their dc polarizability is very small~\cite{Christianen_2019b}. We therefore assume that they undergo ballistic expansion with a velocity distribution corresponding to the temperature of the molecule cloud and are additionally accelerated by gravity. This results in a known probability distribution for the position of a complex at a time $t$ after its formation. By setting $t = \tau_{\mathrm{stick}}$ and calculating the integral of this distribution over the box trap volume, we find $P_t$ and hence also $P_\mathrm{dis}$.

In the case of the incoherent mixture of hyperfine states, we take both $s$- and $p$-wave collisions into account using the MCQDT predictions for loss coefficients from Ref.~\cite{Ospelkaus_2010a}. We then calculate the expected complex lifetime by taking a weighted average over $s$- and $p$-wave collisions with their respective complex lifetimes.

\section{Box trap at zero electric field}
\label{appendix-non-levitated-box}
\begin{figure}
\centering
\includegraphics{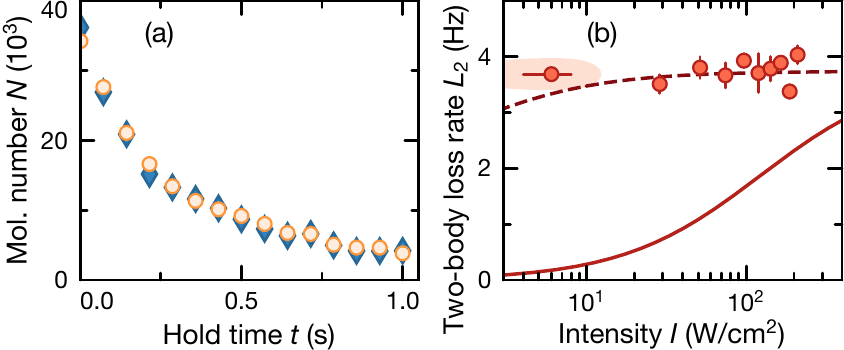}
\caption{Intensity-dependence of loss in the box trap without levitation at $E < \SI{1}{V/cm}$.
(a) Comparison of loss with kill-beam off (blue diamonds) and on at an intensity of $I=\SI{204}{W/cm^2}$ (orange circles). Error bars represent the standard error of the mean of three repetitions. (b) Two-body loss rate $L_2$ for different values of $I$. The solid (dashed) line indicates the expected loss rate assuming a $\tau_{\mathrm{stick}}$ of $\SI{18}{\mu s}$ ($\SI{1.4}{ms}$). The shaded area indicates the 3$\sigma$ confidence region of the lowest-intensity data point.}
\label{fig-loss-non-levitated-box}
\end{figure}
Because the levitation of molecules requires a sizable electric background field, complex decay at very low electric fields can only be studied without levitation. With no voltage applied to the electrodes, we found the residual electric field to be below \SI{1}{V/cm}. Without levitation, molecules accumulate on the bottom of the trap due to gravity, and are closer to the high-intensity walls of the box trap on average. This results in an increased residual light intensity of $\SI{6(2)}{W/cm^2}$ corresponding to $\Gamma_{\mathrm{sc}} = \SI{1.0(3)}{Hz}$. The two-body loss is still dominating due to the larger density. Because we found a precise determination of density to be impractical in this trap configuration, we did not extract $\beta$, but rather the two-body loss rate $L_2 = \beta \bar{n}$, with the mean density $\bar{n}$, from our data, see Fig.\ \ref{fig-loss-non-levitated-box}.

\section{Simple recollision model} 
\label{appendix_recollisions}

\begin{figure}
\centering
\includegraphics{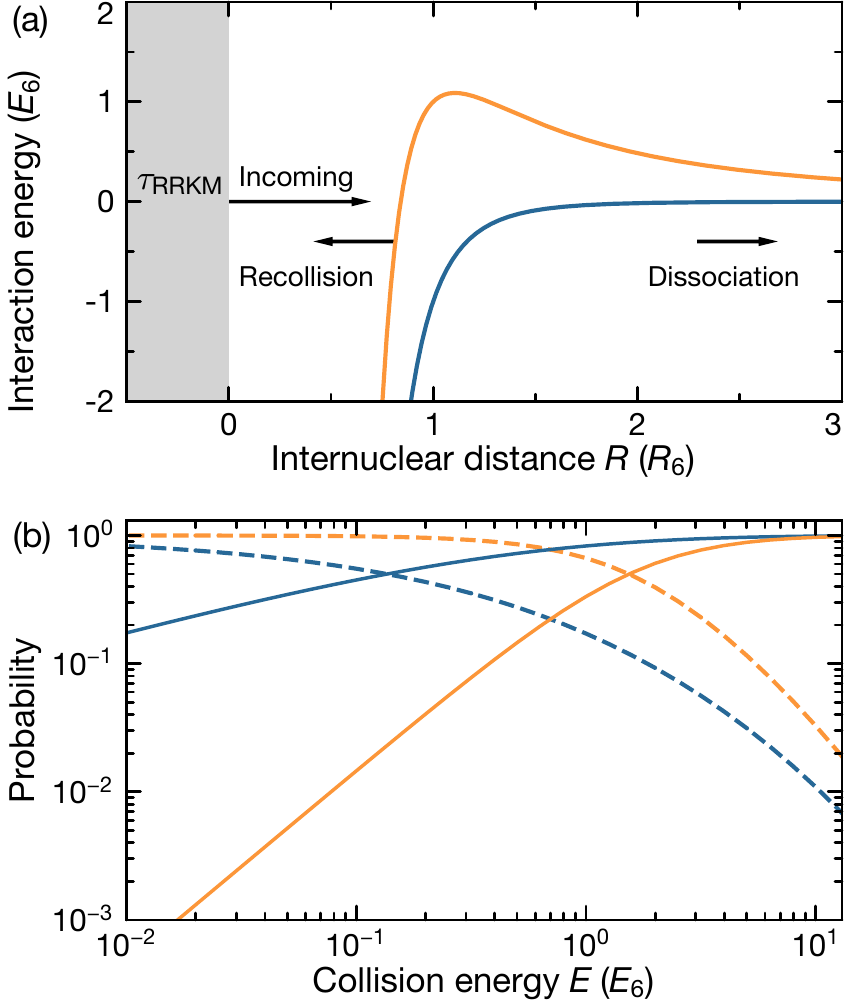}
\caption{ 
Overview of the recollision hypothesis. (a)~Effective radial potential of the van-der-Waals interaction between two molecules for $s$-wave (blue) and $p$-wave (orange) collisions.
When the molecules reach $R=0$, a complex with lifetime $\tau_\mathrm{RRKM}$ is formed.
Dissociation of a complex leads to incoming flux emerging from short range.
Part of this flux is transmitted through the potential which leads to dissociation to two free molecules. However, part of the flux is reflected back to short range, where a short-range complex is again formed. (b)~Reflection (solid lines) and transmission (dashed lines) probabilities at the long-range potential. Probabilities for $s$- and $p$-wave collisions are shown in blue and orange, respectively.}
\label{fig-recollision-overview}
\end{figure}

For dissociation of the short-range complex to yield free molecules,
the molecules must first traverse the long-range potential. Depending on the collision energy, there is however a probability $P_R$ for molecules to be reflected back into the short-range potential.
If the long-range potential contains a barrier, as is the case for collisions of indistinguishable fermions, the probability of successfully traversing the long-range potential, $P_T = 1-P_R$, can be very small.
In the ultracold limit, this probability becomes small even for barrierless long-range potentials.
Hence, molecules may repeatedly form short-range collision complexes, thus enhancing the effective lifetime of the complex as shown in Fig.~\ref{fig-recollision-overview}(a).

To quantify this effect, we performed a single-channel scattering calculation for the Schr\"{o}dinger equation,
\begin{align}
\left( -\frac{d^2}{dr^2} + \frac{L(L+1)}{r^2} - \frac{1}{r^6} - \epsilon\right) \phi(r)=0,
\end{align}
in the reduced units $r=R/R_6$ and $\epsilon=E/E_6$,
where $R_6 = (2 \mu C_6 )^{1/4}$ and $E_6 = 1/(2 \mu R_6^2)$.
The boundary conditions are such that flux is emerging from the short range---due to a short-range complex dissociating---and flux can escape at long range or be absorbed fully when it returns to short range.
We note the probabilities for transmission and reflection from the long-range potential are closely related to the MQDT transmission and reflection amplitudes of Refs.~\cite{Gao_2010, Wang_2012}.
The calculated probabilities are shown in Fig.~\ref{fig-recollision-overview}(b).
The mean number of times the complex forms is $\mathcal{N} = 1/P_T$, such that $\tau_{\mathrm{stick}} = \tau_{\mathrm{RRKM}} \mathcal{N}$.
The effective lifetime of a complex is extended by a factor $\mathcal{N}$,
which can be substantial if $E\ll E_6$, especially for $p$-wave collisions.
For NaK, $E_6 = \SI{11.4}{\mu K}$, such that at a typical experimental temperature of $\SI{0.5}{\mu K}$ we should expect $\mathcal{N}_{s\text{-wave}} = 3$ and $\mathcal{N}_{p\text{-wave}} = 270$.
For RbCs, $E_6 = \SI{3.6}{\mu K}$, so if that experiment is performed around $\SI{2}{\mu K}$ we should expect $\mathcal{N}_{s\text{-wave}} = 1.4$,
such that recollisions do not play an important role in the experiment of Ref.\ \cite{Gregory_2020}. A more rigorous study using MQDT is currently in progress.

\section{Temporally dark dipole trap}
\label{appendix-chopped-trap}
\begin{figure}
\centering
\includegraphics{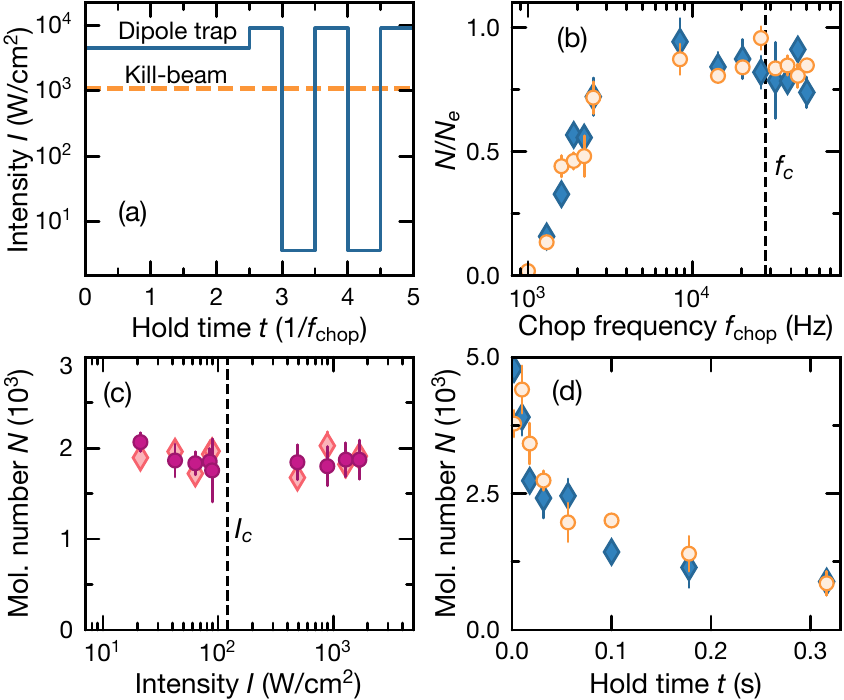}
\caption{Molecules in the chopped crossed dipole trap. (a)~Time-dependent intensity of dipole trap and kill-beam without (left side) and with chopping (right side). (b)~Normalized remaining molecule number versus chopping frequency with kill-beam at $I=\SI{1.08}{kW/cm^2}$ (orange circles) and without kill-beam (blue diamonds). (c)~Dependence of remaining molecule number on kill-beam intensity at chopping frequencies of \SI{2}{kHz} (bright diamonds) and \SI{25}{kHz} (dark circles). The vertical dashed lines indicate the predicted critical values of $f_c = 1/(2\tau_\text{RRKM})$ and $I_{c,1064}$ from Ref.\ \cite{Christianen_2019a}. (d)~Loss of molecules in the excited hyperfine state $|0, 0, -1/2, -4 \rangle$ at $f_{\mathrm{chop}}=\SI{2.5}{kHz}$ with kill-beam at $I=\SI{1.08}{kW/cm^2}$ (orange circles) and without kill-beam (blue diamonds).  All error bars represent the standard error of the mean of 3-5 repetitions.}
\label{fig-loss-chopped-trap}
\end{figure}
To support our conclusions and to exclude the possibility of the box trap causing additional loss in some way, we probed the molecular lifetime in a chopped, red-detuned dipole trap as previously demonstrated by Gregory \textit{et al.~}\cite{Gregory_2020}. The dipole-trap intensity was modulated such that it remains the same on average, but is temporarily reduced significantly. If the dark time is longer than $\tau_{\mathrm{stick}}$, some complexes should dissociate, thus reducing the observed molecule loss. This further requires the residual intensity during the dark time to be on the order of the critical intensity $I_c$, where half of the complexes are expected to dissociate. 

In contrast to the data presented in all other sections of this paper, the data in this section were taken with molecules produced with a different STIRAP method, described in Ref.\ \cite{Seesselberg_2018a}. 
With this method, we initially produced molecules in the $|0, 0, -1/2, -4\rangle$ state, and then used two microwave $\pi$-pulses to transfer them to the hyperfine ground state $|0, 0, 3/2, -4\rangle$. 

The coupling between the two states was realized via a rotationally-excited intermediate state with predominant $|1, -1, 3/2, -4\rangle$ character and some $|1, 1, -1/2, -4\rangle$ admixture at a magnetic field of \SI{85.5}{G}~\cite{Li_2018}. The transition energy of the up-leg (down-leg) was $h \times \SI{5.6432235}{GHz}$ ($h \times \SI{-5.6434105}{GHz}$) and we set the Rabi frequency of the corresponding resonant $\pi$-pulse to $2\pi \times \SI{5.43}{kHz}$ ($2\pi \times \SI{3.23}{kHz}$). The down-leg Rabi frequency was limited by the fact that about $h \times \SI{5}{kHz}$ above the intermediate state there is a state with some $|1, 0, 3/2, -4\rangle$ character, which would also couple to the hyperfine ground state. During the microwave transfer the dipole traps were turned off to avoid broadening of the transition due to inhomogeneous light shifts. About 14\% of the molecules were lost during the state transfer.

For these measurements, the molecules remained in the crossed-beam optical dipole trap. The intensity of the trap beams was chopped using acousto-optical modulators. During the square-wave modulation with 50\% duty cycle, the intensity was doubled during bright times, and reduced to \SI{5}{W/cm^2} (\SI{0.1}{W/cm^2}) for the 1064-nm beam (1550-nm beam) during dark times, see Fig.~\ref{fig-loss-chopped-trap}(a). The dark-time intensities were significantly below the predicted critical intensities $I_{c,1064}$ and $I_{c,1550}$ of \SI{123}{W/cm^2} and \SI{623}{W/cm^2}, respectively~\cite{Christianen_2019a}. Following a holding time in the chopped trap, we reversed the microwave transfer in order to image the molecules.

Initially, we investigated the dependence of the molecule loss on the chopping frequency $f_\text{chop}$. The trap was chopped for $t = \SI{170}{ms}$ (for $f_\text{chop}\leq\SI{2.5}{kHz}$) or \SI{50}{ms} (for $f_\text{chop}>\SI{2.5}{kHz}$). These times were chosen to be long enough that there is already significant molecule number decay after an equally long hold time without chopping. The remaining molecule numbers $N_e$ without chopping are 45\% and 70\% of the initial values, respectively, at the chosen hold times. If the chopping had the desired effect, we should observe that the remaining molecule numbers are increased such that $N/N_e > 1$. For $f_\text{chop}>\SI{2}{kHz}$, we found that the modulation itself causes negligible loss or heating, as shown in Fig.~\ref{fig-loss-chopped-trap}(b). The results were compared to measurements where we added a kill-beam as described previously to ensure that any possible loss caused by the chopping is present in both measurements and can not confound our analysis. If the kill-beam is turned off and $f_\text{chop} < 1/(2\tau_\text{RRKM})$, it is expected that some molecules survive collisions. However, we neither observed suppression of the loss when we chop the trap nor a further enhancement of the loss when we add the kill-beam.

Next, we varied the intensity $I$ of the kill-beam at fixed chopping frequencies, as shown in Fig.~\ref{fig-loss-chopped-trap}(c). For $f_\text{chop} < 1/(2\tau_\text{RRKM})$ we would expect to find a suppression of the molecule loss when $I < I_{c,1064}$. However, after a hold time $t$ of \SI{50}{ms} in the chopped trap, our results are independent of $I$ and consistent with the results in absence of the kill-beam. We also probed the loss of molecules in the state $|0, 0, -1/2, -4\rangle$ in the chopped trap by omitting the microwave transfer and observed no difference between measurements with and without the kill-beam, see Fig.~\ref{fig-loss-chopped-trap}(d).

\bibliography{bibliography}

\end{document}